\documentclass{article}
\usepackage{spconf,amsmath,amsfonts,amssymb,graphicx}
\usepackage[linesnumbered,ruled,vlined]{algorithm2e}
\usepackage{array}
\usepackage{multirow}

\title{Head-synchronous Decoding for Transformer-based Streaming ASR}
\name{Mohan Li, C\u{a}t\u{a}lin Zoril\u{a} and Rama Doddipatla}
\address{Toshiba Cambridge Research Laboratory, Cambridge, UK}
\begin{document}
\ninept
\maketitle
\begin{abstract}
Online Transformer-based automatic speech recognition (ASR) systems have been extensively studied due to the increasing demand for streaming applications. Recently proposed Decoder-end Adaptive Computation Steps (DACS) algorithm for online Transformer ASR was shown to achieve state-of-the-art performance and outperform other existing methods. However, like any other online approach, the DACS-based attention heads in each of the Transformer decoder layers operate independently (or asynchronously) and lead to diverged attending positions. Since DACS employs a truncation threshold to determine the halting position, some of the attention weights are cut off untimely and might impact the stability and precision of decoding. To overcome these issues, here we propose a head-synchronous (HS) version of the DACS algorithm, where the boundary of attention is jointly detected by all the DACS heads in each decoder layer. ASR experiments on Wall Street Journal (WSJ), AIShell-1 and Librispeech show that the proposed method consistently outperforms vanilla DACS and achieves state-of-the-art performance. We will also demonstrate that HS-DACS has reduced decoding cost when compared to vanilla DACS.
\end{abstract}

\section{Introduction}
\label{sec:intro}

Recent advances in neural networks have facilitated to close the gap in ASR performance between End-to-end (E2E) systems and conventional hybrid Hidden Markov Model (HMM) systems. In E2E paradigm, the acoustic model (AM), pronunciation lexicon and language model (LM) that serve as the separate components of the hybrid system, are integrated into a single neural structure, and could be jointly optimised without any prior knowledge. The most popular E2E modelling techniques include: Connectionist Temporal Classification (CTC) \cite{graves2006, graves2014}, Recurrent Neural Network Transducer (RNN-T) \cite{graves2012}, and attention-based encoder-decoder architectures \cite{chorowski2015,chan2016,chiu&sainath2018}.

RNN has been the most popular modeling approach until the advent of Transformer \cite{vaswani2017} architectures. They have been shown to outperform the RNN counterpart on a number of ASR tasks \cite{dong2019,zhao2019,karita2019}. Compared to the sequential processing method used in RNN-based models, Transformer adopts the self-attention mechanism to capture the dependencies between each pair of elements in the input sequence, which effectively breaks the limit of distance and reduces the computational complexity.

Similar to other encoder-decoder architectures, the latency issues encountered in Transformer ASR makes it difficult to deploy in streaming/online applications. The system requires access to full speech utterances before it can perform decoding. To overcome this limitation, online attention mechanisms have been proposed, and can be broadly classed into the following categories: i) \emph{Bernoulli-based attention} methods formulate the triggering of outputs as a stochastic process, and the decision is sampled from the attending probabilities. These include Hard Monotonic Attention (HMA) \cite{luong2017}, Monotonic Chunkwise Attention (MoChA) \cite{chiu2018,miao2019} and Monotonic Truncated Attention (MTA) \cite{miao2020a}; ii) \emph{Triggered attention} methods are conditioned on the forced alignment produced by CTC \cite{moritz2019,moritz2020}; iii) \emph{Accumulation-based attention} methods accumulate the attention weights along encoding timesteps and the computation is halted once the sum reaches a certain threshold. These include Adaptive Computation Steps (ACS) \cite{li2019,dong2020} and Decoder-end Adaptive Computation Steps (DACS) \cite{li2020}.

Among all the aforementioned approaches, the accumulation-based attention mechanism has achieved the state-of-the-art ASR performance with the DACS online Transformer architecture. Identical to other streaming methods, DACS adopts asynchronous multi-head attention mechanism for the Transformer decoder layers, which means the heads operate independently to detect the positions where to attend. Though multiple heads bring more flexibility and robustness to the system, the stability of online decoding could be undermined by the asynchronous working style. Specifically, the diverged behaviours of the heads may lead to ambiguous attention boundaries as it relies on post-processing steps to synchronise. Moreover, using the same but separate truncating threshold for all the heads could cause the loss of attention weights, as some heads are more active than the others. Therefore, in this work we extend and improve the DACS concept by proposing a head-synchronous DACS (HS-DACS) algorithm that is customised for the multi-head attention mechanism employed in Transformer architecture, which assures unified attention boundaries for each decoder layer, and possibly exploits the power of all attention heads.

The rest of the paper is organised as follows: Section 2 presents the architecture of Transformer ASR. Section 3 illustrates the details of the vanilla DACS algorithm. Section 4 presents the proposed head-synchronous version. Experimental results are provided in Section 5. Finally, conclusions are drawn in Section 6.

\section{Transformer-based Online ASR system}
\label{sec:transasr}

The Transformer architecture for ASR systems is similar to the one adopted in neural language processing tasks \cite{vaswani2017}, which consists of a self-attention encoder (SAE) and a self-attention decoder (SAD). In contrast to recurrent modelling methods, Transformer allows each element in a sequence to equally access any other ones regardless of the distance between them, which is achieved by Scaled Dot-Product Attention mechanism:
\begin{equation} \mathrm{Attention}(\mathbf{Q},\mathbf{K},\mathbf{V}) = \mathrm{softmax}(\frac{\mathbf{Q} \mathbf{K}^T} {\sqrt{d_k}}) \mathbf{V}, \label{eq: dotproductatt} \end{equation}
where $\mathbf{Q,K,V} \in \mathbb{R} ^{T/L\times d_k}$ are identical and denote the state matrices of either encoder or decoder in the self-attention sub-layers, whereas $\mathbf{Q} \in \mathbb{R} ^{L\times d_k}$ denotes the decoder states, and $\mathbf{K,V} \in \mathbb{R} ^{T\times d_k}$ are identical as the encoder states in the cross-attention sub-layers, with $d_k$ the dimension of the representation and $T$, $L$ the length of the encoder and decoder states, respectively.

The employment of multiple attention heads further enhances the modelling abilities of Transformer, where $\mathbf{Q,K}$ and $\mathbf{V}$ are projected into diverged sub-spaces and help to compute attentions from different aspects:
\begin{equation} \mathrm{MultiHead}(\mathbf{Q},\mathbf{K},\mathbf{V}) = \mathrm{Concat}(\mathrm{head}_1,...,\mathrm{head}_H) \mathbf{W}^O, \label{eq:multihead} \end{equation}
\begin{equation} \mathrm{where\;head}_h = \mathrm{Attention}(\mathbf{QW}^Q_h,\mathbf{KW}^K_h,\mathbf{VW}^V_h), \label{eq:projectheads} \end{equation}
where $\mathbf{W}^{Q,K,V}_h \in \mathbb{R} ^{d_k \times d_k}$ and $\mathbf{W}^O \in \mathbb{R} ^{d_m \times d_m}$ are the weights of projection layers, and $d_m = H\times d_k$, given $H$ as the number of attention heads.

As reflected in eq. (\ref{eq: dotproductatt}), the attention is computed upon the full sequence of encoder and/or decoder states as required by the softmax function, which poses a big challenge for online recognition. To stream the Transformer ASR system, chunk-hopping based strategies \cite{dong2019,tsunoo2019a,tsunoo2019b,miao2020b} have been applied on the encoder side, where the input utterance is spliced into overlapping chunks and the chunks are chronologically fed to the SAE. Thus, the latency of the online encoder is subject to the chunk size. As for the decoder is concerned, the most popular streaming strategies are the online hard attention mechanisms. Monotonic Chunkwise Attention (MoChA) is one of the first methods applied to Transformer's SAD as in \cite{tsunoo2019b}. During decoding, an attention energy is monotonically calculated for each encoding timestep and passed onto a sigmoid unit to produce an attending probability, from which the decision is sampled to indicate whether to trigger an output. Also, a second-pass soft attention is performed on a small chunk of encoder states that end at the attended timestep, in order to add flexibility to the speech-to-text alignment. Another attempt to enable online decoding is based on Monotonic Truncated Attention (MTA) \cite{miao2020b}, which simplifies MoChA's training criterion and rules out the second-pass soft attention mechanism. 

However, a common drawback of the above Bernoulli-based online attention mechanisms is the difficulty of controlling latency, owing to the distinct behaviours of the multiple heads in Transformer's decoder layers. As some of the heads may produce constant weak attending probabilities, the output could never be triggered for the entire utterance, leading to increasing latency. The recently proposed Decoder-end Adaptive Computation Steps (DACS) algorithm \cite{li2020} circumvents this problem by turning the stochastic process into an a more stable accumulation of attention confidence. For each decoding step, the monotonic attention mechanism is halted either when the accumulation value exceeds a set
threshold or when surpassing a fixed number of timesteps, whichever is earlier. A brief description of the DACS algorithm is presented in the following section.

\section{Decoder-end Adaptive Computation Steps}
\label{ssec:dacs}

The workflow of the DACS algorithm is elaborated below for the Monotonic Attention (MA) $head_h$ in the $l^{th}$ SAD layer for instance. At output step $i$, a MA energy $e_{i,j}$ is computed for each encoding timestep $j$ by the Scaled Dot-Product equation, given the encoder state $k_j$ and decoder state $q_{i-1}$:
\begin{equation} e_{i,j} = \frac {q_{i-1} k_{j}^T} {\sqrt{d_k}}, \end{equation}
which corresponds to the term inside the softmax function of eq. (1) but is generated in the step-wise fashion. The energy is then input into a sigmoid unit to produce an attention weight:
\begin{equation} p_{i,j} = \mathrm{Sigmoid}(e_{i,j}), \end{equation}
which is also known as the \emph{halting probability} that presents the confidence of terminating the computation at the current timestep. The sigmoid function is regarded as an effective alternative to the softmax, which scales down the energy value and precludes the need of global normalisation. From $j=1$, we keep computing and accumulating $p_{i,j}$ monotonically, and halt the computation as soon as the accumulative sum reaches the threshold, which in this case is 1. Meantime, a maximum look-ahead step, $M$, is introduced to force the termination in case that $p_{i,j}$ is constantly small and the accumulation fails to exceed 1 for a fixed number of timesteps. The above process can be summarised by defining an adaptive computation step:
\begin{equation} N_i = \mathrm{min} \Bigg\{ \mathrm{min} \Bigg\{ n: \sum_{j=1}^n p_{i,j} > 1 \Bigg\}, M \Bigg\}, \end{equation}
which denotes the minimum number of encoding timesteps that $head_h$ requires to generate the current output. As a result, a context vector is produced as:
\begin{equation} c_i = \sum_{j=1}^{N_i} p_{i,j} v_{j}, \end{equation}
where $v_j$ equals to the encoder state $k_j$, and the halting probability $p_{i,j}$ directly serves as the attention weight without any normalisation operations like the softmax function.

To take full advantage of the encoding history, after performing the DACS algorithm on all the MA heads in the Transformer SAD, it is necessary to synchronise their halting positions, first within individual layers and then across all layers, so as to reorganise the decoding pace of the whole structure. This is achieved by selecting the furthest encoding timestep reached by any of the heads as the unified halting position for the decoder, and accordingly set the new look-ahead limit for the next output step.

As discussed earlier, in each SAD layer of the DACS-based Transformer system, the multiple MA heads independently perform the adaptive computation steps with an identical threshold of 1. The halting position in DACS is determined either when the accumulation threshold is reached or when the computation exceeds a fixed number of steps. This could lead to the following situation: for most output steps, only a certain part of the heads are active enough to be halted by the threshold, while the rest keep idle and are eventually truncated by the limit of computation steps. Besides, There is a possibility that the halting position for the active heads is not ideal and substantial attention weights are left beyond the boundary by the abrupt halt, giving rise to potential degradation of performance. To overcome the above, the Head-Synchronous Decoder-end Adaptive Computation Steps (HS-DACS) algorithm is proposed, where all the MA heads in a SAD layer contribute simultaneously to the accumulation of halting probabilities, hence are forced to halt at the same position and could flexibly share a joint-threshold during an output step. More details about
the proposed HS-DACS algorithm are presented in the following section.

\section{Proposed Head-synchronous DACS}
\label{ssec:hs-dacs}

To illustrate the workflow of the HS-DACS algorithm, $p_{i,j}$ in eq. (5) is firstly extended to the layer-wise form by aggregating over all the MA heads:
\begin{equation} p^l_{i,j} = \sum_{h=1}^H p^h_{i,j}, \end{equation}
where $l$ denotes the index of the SAD layer that comprises of $H$ heads. Then we again accumulate $p^l_{i,j}$ monotonically from $j=1$, but the computation is now halted when the accumulation is greater than $H$. Similarly, a maximum look-ahead step $M$ is imposed to prevent from reaching the end of speech too fast. As a result, a layer-wise adaptive computation step $N^l_{i}$ is redefined from eq. (6) as:
\begin{equation} N^l_i = \mathrm{min} \Bigg\{ \mathrm{min} \Bigg\{ n: \sum_{j=1}^n p^l_{i,j} > H \Bigg\}, M \Bigg\}. \end{equation}
One direct convenience brought by the HS-DACS algorithm is the spared need of post-synchronisation carried out at each SAD layer, as a sole halting position is naturally produced for each output step. On the other hand, the sum of halting probabilities calculated by each head might no longer be capped at 1. Instead, the active heads could account for more attention confidence that is compromised by the inactive ones, which effectively helps reduce the loss of attention beyond the joint halting position. Such a property is clearly reflected in Figure 1, where one can observe that the MA heads in the vanilla DACS system produce similar and plausible speech-to-text alignments, while the HS-DACS system seems to rely on one dominant head, with the others complementing attention weights from different parts of the input utterance. The pseudo-code of the HS-DACS inference is provided in Algorithm 1.

\begin{figure}[t]

\begin{minipage}[t]{1.0\linewidth}
  \centering
  \centerline{\includegraphics[width=8.5cm]{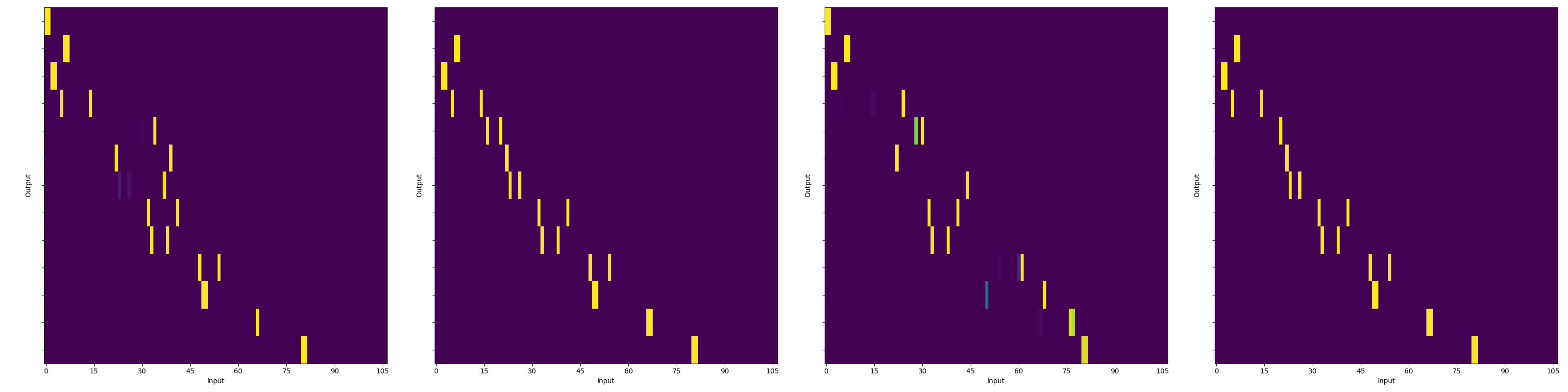}}
  \centerline{(a) DACS}\medskip
\end{minipage}
\begin{minipage}[t]{1.0\linewidth}
  \centering
  \centerline{\includegraphics[width=8.5cm]{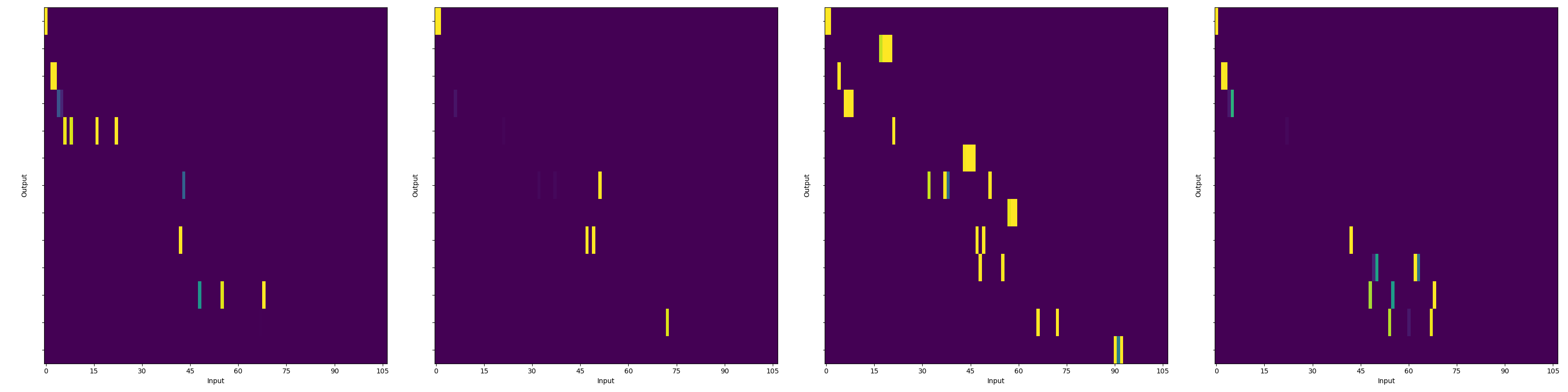}}
  \centerline{(b) HS-DACS}\medskip
\end{minipage}
\caption{Monotonic attention weights produced by the different heads at the top SAD layer of DACS and HS-DACS based Transformer ASR systems. The horizontal axis represents the encoding timestep and the vertical axis represents the output step. The example utterance is chosen from the dev set of AIShell-1 dataset.}
\label{fig:res}
\end{figure}

\begin{algorithm}[t]
\DontPrintSemicolon
\SetAlgoLined
\KwIn{encoder states $\mathbf{k}(\mathbf{v})$, decoder states $\mathbf{q}$, length $T$, maximum look-ahead step $M$, number of heads $H$, number of decoder layers $N_d$. }
 \textbf{Initialization:} $y_0=\langle sos \rangle$, $t_0=0$ \;
 \While{$y_{i-1} \neq \langle eos \rangle$} {
      \For{$l = 1$ \textbf{to} $N_d$} {
      $acc^{h,l}_i=0$ \;
      \For{$j = 1$ \textbf{to} \rm{min}($t_{i-1}+M,T$)} {
        \For{$h = 1$ \textbf{to} $H$} {
            $p^{h,l}_{i,j}=\mathrm{sigmoid}(\frac{q^{h,l}_i k^{T}_j}{\sqrt{d_k}})$ \;
            $acc^{h,l}_i \mathrel{{+}{=}} p^{h,l}_{i,j}$ \;
            }
            \If{$acc^{h,l}_i > H$} {
                break \;
            }
        }
        \For{$h = 1$ \textbf{to} $H$} {
            $c^{h,l}_i=\sum^j_{m=1}p^{h,l}_{i,m}v^{h,l}_{m}$ \;
        }
        $c^l_i=\mathrm{Concat}(c^{1,l}_i,...,c^{H,l}_i)$ \;
        $t_i=\mathrm{max}(t_i,j)$ \;
    }
    $i \mathrel{{+}{=}} 1$ \;
  }
 \caption{HS-DACS Inference for Transformer ASR}
\end{algorithm}

\section{Experiments}
\label{sec:exp}

\subsection{Experimental setup}
\label{ssec:expsetup}

The proposed algorithm has been evaluated on three datasets: Wall Street Journal (WSJ), Librispeech \cite{panayotov2015} for English tasks, and AIShell-1 \cite{bu2017} for a Chinese task, based on the recipes in ESPNET toolkit \cite{watanabe2019}. The acoustic features are composed of 80-dimensional filter-bank coefficients and 3-dimensional pitch information. The numbers of output classes for WSJ and AIShell-1 are 52 and 4231, respectively. And for Librispeech, BPE sub-word tokenisation \cite{rico2016} is applied to construct a vocabulary of 5000 word-pieces.

A similar Transformer architecture is adopted for all the three tasks. The online encoder is similar to the one presented in \cite{miao2020b}, which is a 6-layer chunk-SAE. The sizes of central, left and right chunks are identical and have a dimension of 64. The front-end module consists of 2 CNN layers, with each having 256 kernels of size $3\times3$ and a stride of $2\times2$ that subsamples the input frames by 2-fold. The online decoder is composed of 12 HS-DACS based SAD layers, where the attention dimension, number of heads and size of FFN units are \{256, 4, 2048\} for WSJ and AIShell-1, and \{512,8,2048\} for Librispeech respectively.

We conduct CTC/attention joint training with the CTC weight of 0.3 for all tasks. The models are trained up to 50 epochs for AIShell-1 and 100 epochs for WSJ and Librispeech. An early stopping criterion with a patience of 3 is only imposed on WSJ. The learning rate is scheduled using the Noam weight decay strategy \cite{vaswani2017}, where the initial value is set to 10 for WSJ and Librispeech, and 1 for AIShell-1, with the warm-up steps all set to 25000. Label smoothing and attention dropout are also applied with the factor of 0.1 to overcome the over-fitting problem. Specifically, the maximum look-ahead step M is not employed during training.

During inference, the CTC score is also used as an auxiliary term to combine with Transformer's output distribution, with the weights of \{0.3,0.5,0.4\} for WSJ, AIShell-1 and Librispeech respectively. External LMs are trained as well to rescore the hypothesis produced from the beam-search decoding, which are an 1000-unit 1-layer, a 650-unit 2-layer and a 2048-unit 4-layer Long Short Term Memory (LSTM) network for the tasks in the above order. M is set to 16 for all the SAD layers, corresponding to the latency introduced by the chunk-SAE.

\subsection{Results}
\label{ssec:res}

\makeatletter
\newcommand{\thickhline}{%
    \noalign {\ifnum 0=`}\fi \hrule height 1pt
    \futurelet \reserved@a \@xhline
}
\newcolumntype{"}{@{\hskip\tabcolsep\vrule width 1pt\hskip\tabcolsep}}
\makeatother

\begin{table}[t]
\centering
\caption{Word error rates (WERs) on WSJ.}
\begin{tabular}{llll}
\hline \thickhline
\multicolumn{1}{l}{Model}  & \multicolumn{1}{c}{dev93} & \multicolumn{1}{c}{eval92} \\ \thickhline \hline
\multicolumn{3}{l}{Offline}                                                               \\ \hline
\multicolumn{1}{l}{Transformer \cite{karita2019}}   & \multicolumn{1}{c}{-}      & \multicolumn{1}{c}{4.9}       \\ \thickhline \hline
\multicolumn{3}{l}{Online}  \\ \hline
\multicolumn{1}{l}{MoChA Transformer \cite{tsunoo2019b}}  & \multicolumn{1}{c}{-}      & \multicolumn{1}{c}{6.6}  \\
\multicolumn{1}{l}{DACS Transformer \cite{li2020}}       & \multicolumn{1}{c}{8.9}      & \multicolumn{1}{c}{5.5}       \\
\multicolumn{1}{l}{\bf{HS-DACS Transformer}}       & \multicolumn{1}{c}{8.8}      & \multicolumn{1}{c}{5.4}     \\
\hline \thickhline
\end{tabular}

\bigskip

\centering
\caption{Character error rates (CERs) on AIShell-1.}
\begin{tabular}{llll}
\hline \thickhline
\multicolumn{1}{l}{Model}  &&  \multicolumn{1}{c}{dev} & \multicolumn{1}{c}{test} \\ \thickhline \hline
\multicolumn{3}{l}{Offline}                                                               \\ \hline
\multicolumn{1}{l}{Transformer \cite{karita2019}}    & & \multicolumn{1}{c}{-}      & \multicolumn{1}{c}{6.7}       \\ \thickhline \hline
\multicolumn{3}{l}{Online}      \\ 
\hline
\multicolumn{1}{l}{MMA-MoChA Transformer \cite{inaguma2020}}      & & \multicolumn{1}{c}{-}      & \multicolumn{1}{c}{7.5}       \\
\multicolumn{1}{l}{MoChA Transformer \cite{tsunoo2019b}} && \multicolumn{1}{c}{-}      & \multicolumn{1}{c}{9.7}       \\
\multicolumn{1}{l}{BS-DEC Transformer \cite{tsunoo2020}}    && \multicolumn{1}{c}{6.4}      & \multicolumn{1}{c}{7.3} \\
\multicolumn{1}{l}{DACS Transformer \cite{li2020}}   & & \multicolumn{1}{c}{6.5}      & \multicolumn{1}{c}{7.1} \\
\multicolumn{1}{l}{\bf{HS-DACS Transformer}}  && \multicolumn{1}{c}{6.2}  & \multicolumn{1}{c}{6.8}\\ 
\hline \thickhline
\end{tabular}

\bigskip

\centering
\caption{Word error rates (WERs) on Librispeech.}
\begin{tabular}{lllll}
\hline \thickhline
\multirow{2}{*}{Model}      & \multicolumn{2}{c}{dev}   & \multicolumn{2}{c}{test}     \\ \cline{2-5} 
                        & \multicolumn{1}{c}{clean}    & \multicolumn{1}{c}{other} & \multicolumn{1}{c}{clean}    & \multicolumn{1}{c}{other} \\ \thickhline \hline
\multicolumn{3}{l}{Offline}  \\ \hline
\multicolumn{1}{l}{Transformer (ours)}  & \multicolumn{1}{c}{2.4}  & \multicolumn{1}{c}{6.0} 
                                        & \multicolumn{1}{c}{2.6}  & \multicolumn{1}{c}{6.1}    \\ \thickhline \hline
\multicolumn{3}{l}{Online}                                                                \\ \hline
\multicolumn{1}{l}{Triggered attention \cite{moritz2020}}    & \multicolumn{1}{c}{-}  & \multicolumn{1}{c}{-} 
                                        & \multicolumn{1}{c}{2.8}  & \multicolumn{1}{c}{7.2} \\
\multicolumn{1}{l}{CIF \cite{dong2020}}    & \multicolumn{1}{c}{-}  & \multicolumn{1}{c}{-} 
                              & \multicolumn{1}{c}{3.3}  & \multicolumn{1}{c}{9.6} \\
\multicolumn{1}{l}{BS-DEC Transformer \cite{tsunoo2020}}    & \multicolumn{1}{c}{2.5}  & \multicolumn{1}{c}{6.8} 
                                        & \multicolumn{1}{c}{2.7}  & \multicolumn{1}{c}{7.1} \\
\multicolumn{1}{l}{DACS Transformer}    & \multicolumn{1}{c}{2.5}  & \multicolumn{1}{c}{6.6} 
                                        & \multicolumn{1}{c}{2.7}  & \multicolumn{1}{c}{6.8} \\
\multicolumn{1}{l}{\bf{HS-DACS Transformer}}  & \multicolumn{1}{c}{2.4}  & \multicolumn{1}{c}{6.5}
                                               & \multicolumn{1}{c}{2.7}  & \multicolumn{1}{c}{6.6} \\ 
\hline \thickhline
\end{tabular}
\end{table}

\begin{table}[t]
\centering
\caption{Comparison of the decoding cost between DACS and HS-DACS based systems on the test set of AIShell-1. The ratio is averaged over the utterances of the test set.}
\begin{tabular}{lllp{0.1cm}lll} 
\hline \thickhline
\multicolumn{3}{c}{DACS} &  & \multicolumn{3}{c}{HS-DACS}  \\ 
\cline{1-3} \cline{5-7} 
\multicolumn{1}{c}{thr}    & \multicolumn{1}{c}{CER(\%)}    & \multicolumn{1}{c}{$r$} &
& \multicolumn{1}{c}{joint-thr}    & \multicolumn{1}{c}{CER(\%)}    & \multicolumn{1}{c}{$r$} \\ 
\thickhline
\multicolumn{1}{l}{1.0}    & \multicolumn{1}{c}{7.1}  & \multicolumn{1}{c}{0.63} & 
    & \multicolumn{1}{c}{4.0} & \multicolumn{1}{c}{6.8} & \multicolumn{1}{c}{0.60} \\
\multicolumn{1}{l}{0.75}   & \multicolumn{1}{c}{7.2}  & \multicolumn{1}{c}{0.58} & 
    & \multicolumn{1}{c}{3.0} & \multicolumn{1}{c}{6.9} & \multicolumn{1}{c}{0.55} \\
\multicolumn{1}{l}{0.5}    & \multicolumn{1}{c}{7.3}  & \multicolumn{1}{c}{0.58} & 
    & \multicolumn{1}{c}{2.0} & \multicolumn{1}{c}{7.0} & \multicolumn{1}{c}{0.49} \\
\multicolumn{1}{l}{0.25}   & \multicolumn{1}{c}{7.3}  & \multicolumn{1}{c}{0.58} & 
    & \multicolumn{1}{c}{1.0} & \multicolumn{1}{c}{7.9} & \multicolumn{1}{c}{0.40} \\
\thickhline
\end{tabular}
\end{table}


Table 1, 2 and 3 present the results of the proposed method on WSJ, AIShell-1 and Librispeech data respectively. The performance of the proposed method is compared against the performance of vanilla  DACS and other online Transformer based approaches proposed in literature. One can observe that the HS-DACS has significant gains in performance when compared against Bernoulli-based or Triggered-attention based methods. The HS-DACS also consistently outperforms vanilla DACS on all the tasks. For WSJ, we observe that HS-DACS has a relative gain of 18.2\% WER when compared with MoChA Transformer in \cite{tsunoo2019b}. And for AIShell-1 and Librispeech, HS-DACS has a relative gain of 6.8\% WER and 7.0\% WER when compared with BS-DEC Transformer presented in \cite{tsunoo2020} respectively. To our best knowledge, the proposed system has achieved the state-of-the-art online ASR performance on all three tasks.

\subsection{Decoding Cost Comparison}
\label{ssec:latency}

Table 4 compares the decoding cost between DACS and HS-DACS based Transformer ASR systems reported on AIShell-1. Similar to \cite{li2020}, this is measured in terms of the computation cost committed by the MA heads in the cross-attention sub-layers. A computation-step-coverage ratio is therefore defined as:
\begin{equation}
r = \frac{\sum^{N_d}_{l=1} \sum^{H}_{h=1} \sum^{L}_{i=1} {s^{h,l}_i} }{(N_d \times H \times L) \times T},
\end{equation}
where $s^{h,l}_i$ in the numerator denotes the adaptive computation steps consumed by $head_h$ in the $l^{th}$ SAD layer at output step $i$, and is summed up for all $L$ output steps, $H$ heads and $N_d$ SAD layers. Correspondingly, the number of encoding timesteps $T$ is aggregated in the same way by the denominator. The lower the ratio is, the smaller the cost is present during decoding. On the other hand, in our experiments, it is noticed that the maximum look-ahead step is triggered for a majority of output steps, which means the joint-threshold used in the HS-DACS algorithm couldn't be reached by the accumulation of attention confidence. Therefore, we investigate reducing the joint-threshold in order to cut down the decoding cost further, while observing the effect on the ASR performance.

As shown in Table 4, we gradually decay the joint-threshold of the HS-DACS algorithm from 4.0 to 1.0, and meanwhile proportionally reduce the independent threshold of the vanilla DACS algorithm from 1.0 to 0.25 for a fair comparison. It can be seen that the HS-DACS system yields constantly smaller decoding cost than the DACS one for all the threshold settings, and that the ASR result remains stable until the joint-threshold goes down to 2.0. What also worth noticing is that the CER degradation of the DACS system is rather neglectable even when the threshold drops to 0.25, compared with the 1.1\% absolute CER increase seen by the HS-DACS system. We believe that this is because the attention weights tend to be concentrated within a single head of the HS-DACS layer, thus the performance is much more sensitive to the change of threshold than DACS layers, where attentions are evenly distributed across the heads.

\section{Conclusion}
\label{sec:majhead}

The paper presented an approach to perform head-synchronous decoding by improving the recently proposed DACS algorithm for online Transformer ASR. The proposed HS-DACS algorithm drives all the MA heads to jointly compute the adaptive computation step to determine the halting probability, which in turn facilitated to synchronise all the heads for each decoder layer during decoding. This modification eliminated the need for an additional step to post-synchronise the heads in DACS and further helped reduce the cost in decoding. The HS-DACS might have helped mitigate the impact, if any, due to abrupt halting introduced by either the accumulation threshold or a fixed maximum look-ahead step in DACS. We showed through ASR experiments that the proposed HS-DACS consistently outperformed DACS on various tasks and achieved state-of-the-art performance, which is 5.4\% WER, 6.8\% CER and 2.7\%/6.6\% WER on the test sets of WSJ, AIShell-1 and Librispeech, respectively.

\newpage

\bibliographystyle{IEEE}
\bibliography{refs}

\end{document}